\documentclass[fontsize=11pt]{scrartcl}	 

\usepackage[english]{babel} 
\usepackage{amsmath,amsfonts,amsthm} 
\usepackage[svgnames]{xcolor} 
\usepackage[hang, small,labelfont=bf,up,textfont=it,up]{caption} 
\usepackage{booktabs} 
\usepackage{fix-cm}	 

\usepackage{amssymb}
\usepackage{slashed}
\usepackage{epsfig}
\usepackage{graphics}
\usepackage{epstopdf}

\usepackage{sectsty} 

\usepackage{fancyhdr} 
\usepackage{lastpage} 

\def\lsim{\raise0.3ex\hbox{$\;<$\kern-0.75em\raise-1.1ex\hbox{$\sim\;$}}}
\def\gsim{\raise0.3ex\hbox{$\;>$\kern-0.75em\raise-1.1ex\hbox{$\sim\;$}}}
\def\etal{{\it et al.}}

\newcommand{\be}{\begin{eqnarray}}
\newcommand{\ee}{\end{eqnarray}}

\newcommand{\mbb}{\mathbb}
\newcommand{\mc}{\mathcal}
\newcommand{\n}{\nonumber\\}

\def\bea{\begin{eqnarray}}
\def\eea{\end{eqnarray}}

\usepackage{titling} 

\title{\bf\textsc Aspects of moduli stabilization  in type IIB string theory} 

\author{Shaaban Khalil, Ahmad Moursy, and Ali Nassar \\[4pt]
Center for Fundamental Physics, Zewail City of Science and Technology,\\[4pt]
12588 Giza, Egypt}

\begin{document}

\maketitle 

\thispagestyle{fancy} 

We review moduli stabilization in type IIB string theory compactification with fluxes. We focus on the KKLT and Large Volume Scenario (LVS).
 We show that the predicted soft SUSY breaking terms in KKLT model are not phenomenological viable. In LVS, the following result for scalar mass, gaugino mass, and trilinear term is obtained: $m_0 =m_{1/2}= - A_0=m_{3/2}$, which may account for Higgs mass limit if $m_{3/2} \sim {\cal O}(1.5)$ TeV.  However, in this case the relic abundance of the lightest  neutralino can not be consistent with the measured limits. We also study the cosmological consequences of moduli stabilization in both models. In particular,  the associated inflation models such as racetrack inflation and K\"ahler inflation are analyzed. Finally the problem of moduli destabilization
and the effect of string moduli backreaction on the inflation models are discussed.



\section*{Introduction}

Ever since the invention of the Kaluza-Klein mechanism it was realized that extra-dimensional models are plagued with massless scalar fields when compactified to 4D. In the original Kaluza-Klein construction, the radius of the $x^5$-circle, $R(x^\mu)$, is not fixed by the dynamics and appears as a scalar field with no potential in the effective 4D theory. This is a generic feature of most compactifications of higher-dimensional gravitational/Yang-Mills theories. The parameters which describe the shape and size of the compactification manifold give rise to massless scalar \textit{moduli} with no potential in 4D (flat directions) i.e., non-stabilized moduli fields. These moduli are gravitationally coupled and as such there existence would be in conflict with experiment (see \cite{Dolgov:1999gk} for a review).

String theory being a candidate for a unified theory of the forces of nature must be able to reproduce the physics of our real world which is 4 dimensional with a small positive cosmological constant and chiral gauge interactions. This amounts to finding de Sitter vacua of string theory with all moduli stabilized. The seminal work in \cite{Candelas:1985en} constructed the first models of Grand Unified Theories (GUTs) from Calabi-Yau compactification of the heterotic string. In these models, the 10D gauge group ($E_8\times E_8$ or $SO(32)$) is broken to Standard-Model-like gauge groups by turning on background gauge fields on the internal space. The chiral fermions are obtained from the dimensional reduction of the 10D gaugino and the number of the generations is half the Euler characteristic of the internal space \cite{Candelas:1985en}. These compactifications are purely geometric and give rise to an $\mathcal{N}=1$ supersymmetric (SUSY) 4D theories with large number of moduli. Other compactifications which leads to  $\mathcal{N}=1$ SUSY in 4D are type II theories on Calabi-Yau orientifolds with D-branes and fluxes, M-theory on manifolds with $G_2$ holonomy, F-theory on Calabi-Yau four-folds (see \cite{Denef:2008wq} for a review and references therein).

One essential fact about our universe is that it has a small positive cosmological constant, i.e., a De Sitter space-time. The moduli scalar potential $V(\phi)$ should be such that it's minimum produces the observed value of the cosmological constant.
This turns out to be a very difficult problem since de Sitter vacua are known to break supersymmetry. The first attempt at finding realistic vacua of string theory were mainly concerned with Minkowski and Anti de Sitter vacua \cite{Candelas:1985en,Strominger:1985it,Strominger:1986uh,Witten:1985bz}. There it was shown that turning on magnetic fluxes in the internal manifold leads to a non-trivial warp factor and a non-K\"ahler geometry.  The situation improved drastically since the introduction of D-branes as non-perturbative objects in string theory. This resulted in the celebrated  KKLT scenario in which a moduli fixing mechanism was introduced  \cite{Kachru:2003aw}. The Large Volume Scenario followed after \cite{Balasubramanian:2005zx}.
Therein one can turn on a vacuum expectation value for the fluxes in the internal space without breaking the 4D Lorentz invariance.



In this paper we  review the flux compactification and moduli stabilization in type IIB string theory. The Kachru-Kallosh-Linde-Trivedi (KKLT) \cite{Kachru:2003aw} scenario was a major step in constructing  dS vacua with all the moduli stabilized including the volume modulus. The Large Volume Scenario (LVS) by Quevedo \etal~ \cite{Balasubramanian:2005zx, Conlon:2005ki} was proposed as an alternative to stabilize all the moduli with the volume moduli stabilized at extremely large volume. Realistic models of moduli stabilization must come as close as possible to the observed phenomenology at low energy and also to account for cosmological inflation at high energy scales. From this point of view, we analyze the low energy phenomenology of both KKLT and LVS. We emphasize that these models provide specific set of soft SUSY breaking terms, which are not  phenomenologically viable.  In addition, we study the impact of these scenarios on inflation. It turns out that inflation could destabilize the moduli again. This problem has been analyzed in details in Ref  \cite{Kallosh:2004yh}. On the other hand moduli stabilization may have backreaction effects on the inflationary potential, which could change the inflationary parameters.

This paper is organized as follows: In Section 1, we review the flux compactification and moduli stabilization in type IIB string theory.  Sections 2 is devoted for KKLT model and its variants, where we discussed several examples for vacuum uplifting.  In Section 3 we present the LVS as an alternative scenario that overcomes some of the KKLT drawbacks. The phenomenological implications and SUSY breaking soft terms of these two models are studied in Section 4. In Section 5 we highlight the  cosmological implications of these two models. Finally we state our conclusions in Section 6.


%
\section{Flux compactifications and moduli stabilization in string theory}\label{sec:IIB}
%
We start by reviewing the heterotic string compactification which has been considered to connect string theory to four dimensions physics \cite{Candelas:1985en}.
Then we discuss type IIB flux compactifications, where the complex structure moduli (CSM) and the dilaton are stabilized by the RR and NS-NS 3-form fluxes.
\subsection{Heterotic string compactification}
\label{ssec:het}
To compactify string theory down to 4D, one looks for vacuum solutions of the form $M_{10}=M_4\times M_6$, where $M_4$ is assumed to have 4D Poincar\'e invariance and $M_6$ (or simply $\cal M$) is a compact internal 6D Euclidean space.
  The most general  metric  compatible with  these requirements can be written as \cite{Candelas:1985en}
  \begin{equation}
  G_{MN}=
  \begin{pmatrix}
  e^{A(y)} \eta_{\mu\nu}&0\\
  0 &g_{mn}(y)
  \end{pmatrix},
  \end{equation}
   where $y^m$ are the coordinates on $\cal M$ with the metric $g_{mn}(y)$ and the requirement of Poincar\'e symmetry of $M_4$ still allows for a warp factor which depends on $\cal M$ only.

   We also require an $\mathcal{N}=1$ supersymmetry in 4D since this is phenomenologically appealing and gives more analytic control. For $M_4$, one consider a homogenous and isotropic maximally symmetric solutions which implies that the Riemann tensor takes the form
\begin{equation}
R_{\mu\nu\rho \lambda} = c \big(g_{\mu\rho} g_{\nu\lambda}-g_{\mu\lambda} g_{\nu\rho}\big),
\end{equation}
  where $c$ is fixed  by contracting both sides with $g^{\mu\rho} g^{\nu\lambda}$ and turns out to be equal to $R/12$. The constant scalar curvature $R$ could be $R=0$ (Minkowski), $R<0$ (AdS), or $R>0$ (dS).

When the radius of curvature of $\cal M$ is large compared to the Planck scale, one can use the supergravity approximation of string theory. In order to have a supersymmetric background, the supergravity transformations  of the fermions must vanish \cite{Candelas:1985en}
\begin{equation}
\delta_\varepsilon (\text{Fermions} )=0.
\end{equation}
For Heterotic string theory, the supergravity variations of the fermions are given by \cite{Candelas:1985en}
\begin{equation}
\begin{split}
\delta \psi_M& = \nabla_M \varepsilon-\frac{1}{4} \mathbf{H}_M \varepsilon ,\\[3pt]
\delta \lambda &= -\frac{1}{2} \slashed \partial \Phi \varepsilon+\frac{1}{4} \mathbf{H} \varepsilon ,\\[3pt]
\delta \xi &= -\frac{1}{2} \mathbf{F} \varepsilon, \\[3pt]
\end{split}
\end{equation}
where $\psi_M, \lambda, \xi$ are the gravitino, dilatino, and the gaugino respectively. The $H$-flux is $\mathbf{H}_M= \gamma_{NP} H_{MNP}$ and $\mathbf{H}= \gamma_{MNP} H_{MNP}$; and $\mathbf{F}=\gamma_{MN} F_{MN}$ is Yang-Mills field strength of the $E_8\times E_8$ or $SO(32)$ gauge fields in 10D.

The Bianchi identity of $H$ is given by
\begin{equation}
dH =\frac{\alpha'}{4}\big[\text{Tr} (R \wedge R )- \text{Tr} (F \wedge F )\big].
\end{equation}
Since $dH$ is exact, i.e., zero in cohomology, then the cohomology classes of $\text{Tr} (R \wedge R )$ and  $\text{Tr} (F \wedge F )$ are the same.

The vanishing of above variations for a given spinor $\varepsilon$ will put some restrictions on the background fields and in particular on the geometry and topology of $\cal M$.
The compactification of heterotic string theory with vanishing $H$-flux (or vanishing torsion)  was first studied in \cite{Candelas:1985en,Strominger:1986uh} and it leads to the following conditions on the string background.
\begin{equation}
\delta \psi_\mu =0 \longrightarrow G_{\mu\nu}=\eta_{\mu\nu}, \quad e^{A(y)}=\text{constant}.
\end{equation}
That is the external space $M_4$ is Minkowski with a constant warp factor.

 The dilatino variation gives
\begin{equation}
\partial_m\Phi =0,
\end{equation}
i.e., the dilaton is constant over $\cal M$.

The gravitino variation $\delta \psi_m$ gives
\begin{equation}
\nabla_m \varepsilon =0.
\end{equation}
This equation says that $\cal M$ admits a covariantly constant spinor. The integrability condition resulting from the above equation implies that $\cal M$ is Ricci flat
\begin{equation}
R_{mn}=0.
\end{equation}
Hence the first Chern class of $\cal M$ vanishes
\begin{equation}
c_1 = \frac{1}{2 \pi} [\mathcal{R}].
\end{equation}
It was conjectured by Calabi and proved by Yau that Ricci-flat compact K\"ahler manifolds with $c_1=0$ admit a metric with $SU(3)$ holonomy. These metrics come in families and are parameterized by continuous parameters $T_i$ which defines the shape and sizes of $\cal M$. The parameters $T_i$ appear as scalar fields (moduli) in 4D with no potential and a major goal in string theory is to generate a potential which stabilize these moduli in a way  consistent with observations.

One can describe the  4D $\mathcal{N}=1$ models resulting from the heterotic string compactification in terms of an effective SUSY theory. This theory is characterized by a K\"ahler potential $K$, a gauge kinetic function $f$, and a superpotential $W$.  The tree-level superpotential $W$ doesn't fix the moduli; due to non-renormalization theorems $W$ is not renormalized at any order in perturbation theory \cite{Seiberg:1993vc,Dine:1986vd}. This means that if supersymmetry is unbroken at tree level, it will remain unbroken to all orders of perturbation theory. Non-perturbative effects such as gaugino condensation \cite{Dine:1985rz} can correct the superpotential and fix some of the moduli.

\subsection{Type IIB compactification}
\label{ssec:IIB}
We now turn our discussion to  type IIB string theory.
The massless bosonic spectrum of type IIB consists of the metric $g_{MN}$,
the RR 0-form, $C_0$ and the scalar dilaton $\phi$ which are combined into the axiodilaton $S = C_0 + i e^{-\phi}$ where the string coupling is given by $\frac{1}{g_s}= e^{-\phi}$.
In addition, the spectrum contains the RR 2-form $C_2$  and 4-form $C_4$ as well as the NS 2-form $B_2$. It is convenient to combine the RR and NS 3-forms $F_3=dC_2$ and $H_3 = dB_2$ into $G_3 = F_3 -  S H_3$.
The classical action of type IIB supergravity $S_{IIB}^0$ is divided into a bulk action $S_b$, the Chern-Simons action $S_{cs}$
and contributions from the D-brane sources $S_l$ \cite{Polchinski} \footnote{More precisely, $S_{l}$ represents the action of the localized sources for the case of a D$p$-brane wrapping a ($p-3$)-cycle $\Sigma$.
}
\bea
\label{SIIB}
S_{IIB}^0 = S_b + S_{cs} + S_l.
\eea
In the string frame, $S_b, S_{cs}$ and $ S_{l}$ are given  by
\bea
\label{iibaction}
S_{b} & = & \frac{1}{(2 \pi)^7 \alpha'^4} \int d^{10}x \sqrt{-g}
\left\{ e^{-2 \phi}[\mc{R} + 4(\nabla \phi)^2] - \frac{F_1^2}{2} -
  \frac{1}{2 \cdot 3!} G_3 \cdot \bar{G}_3 - \frac{\tilde{F}_5^2}{4
    \cdot 5!} \right\},
  \nonumber \\
S_{cs} & = & \frac{1}{4 i (2 \pi)^7 \alpha'^4} \int e^\phi C_4 \wedge
G_3 \wedge \bar{G}_3, \nonumber \\
S_{l} & = & \sum_{sources} \left( - \int_{\mathbb{R}^4\times \Sigma} d^{p+1} \xi  T_p  e^{-\phi} \sqrt{-g}
 + \mu_p \int_{\mathbb{R}^4\times \Sigma} C_{p+1} \right),
\eea
where $T_p$ and $\mu_p$ are respectively, the tension and charge of the D$p$-brane.
The string tension is expressed in terms of string length as
\be
\alpha'=1/M_{st}^2 =(l_s/2\pi)^2.
\ee

The 5-form  ${\widetilde F_5}$ is defined as
\be
{\widetilde F_5} = dC_4 - \frac{1}{2} C_2 \wedge H_3 + \frac{1}{2} F_3 \wedge B_2 ,
\ee
which is self dual and satisfies Bianchi identity
\begin{equation}
d\widetilde F_5 = H_3 \wedge F_3.
\end{equation}

One would like to consider warped compactifications of type IIB on a compact manifold $\cal M$.  The metric ansatz for a 4D warped compactification is given by \cite{Giddings:2001yu}
\begin{equation}\label{wrapedmetric}
ds_{10}^2=\sum_{M,N=0}^9 G_{MN}dx^Md x^N =e^{2A(y)} \eta_{\mu\nu} dx^{\mu} dx^{\nu}+e^{-2A(y)} g_{mn} dy^{m} dy^{n}.
\end{equation}
The 10D Einstein equation of motion is
\begin{equation}
R_{MN}=\kappa^2 \bigg(T_{MN}-\frac{1}{8}G_{MN} T \bigg),
\end{equation}
where
\begin{equation}
T_{MN}=T_{MN}^{\text{sugra}}+T_{MN}^{l},
\end{equation}
is the total stress tensor of supergravity plus the localized objects, {\it i.e}
\begin{equation}
T_{MN}^{l}=-\frac{2}{\sqrt{-G}} \frac{\delta S_{l}}{\delta G^{MN}},
\end{equation}
%

In this regard the space--time components of the latter action reduces to
\begin{equation}
\nabla^2 e^{4A} =\frac{e^{2A}}{12 \text{Im} \tau} |G_3|^2+e^{-6A}\big(|\partial \alpha |^2+\big|\partial e^{4A} \big|^2\big)+\frac{\kappa^2_{10}}{2} e^{2A}\big(T_m^m-T_\mu^\mu\big)^{l},
\end{equation}
where $\alpha$ is a function on the compact space.
Integrating both sides of this equation over the compact manifold $\cal M$, the left-hand side gives zero since it's a total derivative. If there are no localized sources then the right-hand side is a sum of positive terms and vanishes only if $\alpha$ and $A$ are constants and $G_3=0$. This is the familiar no-go theorem of flux compactifications \cite{Maldacena:2000mw, deWit:1986xg}. However, the existence of localized sources in string theory like orientifold planes can balance the contribution coming from fluxes to give a non-trivial warp factor. This was realized in string theory in \cite{Giddings:2001yu}.
The  setup in \cite{Giddings:2001yu} allows for a stabilization of complex structure moduli by turning on RR and NS fluxes in the internal space \cite{Douglas:2006es,Grana:2005jc}.


\subsection{Type IIB fluxes and moduli stabilization}
\label{ssec:IIB-mod-stab}
One way to see the problem in a simple setting is nicely reviewed in \cite{Denef:2007pq,Douglas:2006es} in a toy model and we review it here. One can generate a potential for the moduli by turning on fluxes in the internal space. The potential in 4D results from the Maxwell term of the fluxes
\begin{equation}
V=\int_{\cal M} \sqrt{-g} g^{\mu_1 \nu_1} \cdots g^{\mu_p \nu_p} F_{\mu_1\cdots \mu_p } F_{\nu_1\cdots \nu_p}=\int F_p\wedge * F_p ,
\end{equation}
where $g$ is the metric of the internal space. The metric $g$ will depend on the moduli of the internal space and after doing the integral on $\cal M$, one gets a potential for the moduli $V(\phi)$. For example, consider a 6D Maxwell-Einstein theory compactified on a two sphere $S^2$ with a non-zero flux of $F_2$ piercing $S^2$
\begin{equation}
\int_{S^2} F=N.
\end{equation}
This flux contributes a positive energy to the effective 4D potential which can then balance the negative contribution coming from the curvature of $S^2$.
More specifically the contribution to the effective potential coming from the flux originates from the Maxwell term in 6D
\begin{equation}
V(R)=\int F_2\wedge * F_2\sim \frac{N^2}{R^6},
\end{equation}
where the determinant of the metric contributes a factor of $R^2$ and two metric contractions contribute a factor of $1/R^2$ while the  transformation to the Einstein frame gives a factor $1/R^4$.
Therefor total 4D potential takes the form
\begin{equation}
V(R)=\frac{N^2}{R^6}-\frac{1}{R^4}.
\end{equation}
which is minimized at $R=N$ and if $N$ is large the curvature is small and the supergravity approximation is reliable \cite{Denef:2007pq,Douglas:2006es}. In string theory, additional ingredients beside the fluxes are needed to construct stable vacua. These ingredients are the D-branes and orientifold planes \cite{Blumenhagen:2006ci}.

The main idea of flux compactification is that there are solutions of the string tree-level equations in which some of the $p$-form fields are non-zero in the vacuum. In these constructions, one needs to make sure that the backreaction of the flux on the geometry doesn't take us outside the supergravity approximation. This turns out to be possible \cite{Strominger:1986uh} with the introduction of a warp factor varying over the internal manifold and hence the new geometry is conformal to the non-flux case. The fluxes which can be turned on, are the RR fluxes of type II  and the  $H_3$ flux. In this case the quantization condition on the fluxes is
\begin{equation}
\frac{1}{l_s^p} \int_{\Sigma_{p+1}} F_{p+1} \in \mathbb{Z},
\end{equation}
where the integrality of the cohomolgy classes of $F_{p+1}$ is due to Dirac's charge quantization. The non-vanishing of the cohomology classes of these $p$-form fields leads to obstructions which lifts some of the flat directions of the compactification, i.e., it leads to potential which freezes some of the moduli.


In the presence of sources the modified Bianchi identity now reads \cite{Blumenhagen:2006ci}
\begin{equation}
d\widetilde{F}_5=H_3\wedge F_3 +2 \kappa^2_{10}\mu_3 \sum_a \pi_6^a+
2 \kappa^2_{10} Q_3\mu_3 \pi_6^{O3}.
\end{equation}
Integrating this equation over the compact internal manifold one gets the tadpole cancelation condition \cite{Blumenhagen:2006ci}
\begin{equation}\label{tadpole1}
N_{\text{flux}}+ N_{\text{D}3}+Q_3 N_{O3}=0 ,
\end{equation}
where
\begin{equation}
N_{\text{flux}}=\frac{1}{l_s^4}\int_{\cal M} H_3\wedge F_3.
\end{equation}

The type IIB string theory will be compactified on Calabi-Yau orientifolds in order to obtain a 4D ${\cal N}=1$  model.
 It turned out that one needs to make an orientifold projection in order to have supersymmetric compactification \cite{Blumenhagen:2006ci}. The orientifold action projects out one of the two gravitinos and breaks the $\mathcal{N}=2$ SUSY down to $\mathcal{N}=1$. The orientifold projection  also introduces O-planes with a background charge and a negative energy density which balances the contribution of D-branes and leads to stable compactification.
 The RR and NS-NS 3-form fluxes are restricted via the integral cohomology which determines the quantized background fluxes as follows
\be
\frac{1}{(2 \pi)^2 \alpha'} \int_{\Sigma_a} F_3 = n_a \in \mbb{Z}, \quad \quad \frac{1}{(2 \pi)^2 \alpha'}
\int_{\Sigma_b} H_3 = m_b \in \mbb{Z},
\ee
with  $\Sigma_{a,b}$ are  3-cycles of the Calabi-Yau manifold. In this case tadpole condition (\ref{tadpole1}) reads
\be
{\chi({\cal M})\over 24} ~=~N_{D3} + {1\over {(2 \pi^4) \alpha'^2}}
\int_{\cal M} H_{3} \wedge F_3~,
\label{tadpole}
\ee
where $N_{D3}$ is the
net number of ($D3 - {\overline{D3}}$) branes filling the noncompact dimensions and $\chi({\cal M})$ is the Euler characteristic of the  elliptically fibred  Calabi-Yau fourfold ${\cal M}$.

The compactification to 4D on a Calabi-Yau manifold with orientifold planes will give rise to an $\mathcal{N}=1$ supergravity theory which is characterized by a K\"ahler potential $K$, a superpotential $W$ and a gauge kinetic function $f$ \cite{Wess:1992cp, Ibanez:2012zz}. The tree-level K\"ahler potential is given by the Weil-Petersson
metric using the Kaluza-Klein reduction of type IIB supergravity.
\be
K ~=~-3 \ln[(T + \bar T)] - \ln[(S + \bar S)] -
\ln[-i\int_{\cal M} \Omega \wedge {\overline{\Omega}}].
\label{kahler1}
\ee
Here $T$ represents the volume modulus and is one of the   K\"ahler moduli.
The conditions on the fluxes in type IIB are
\begin{equation}
*G_3=iG_3, \qquad  G_3^{0,3}=0.
\end{equation}
Since the hodge $*$ depends on the metric then one expects  the above conditions can fix the geometric moduli except for the overall scale of the metric since the hodge $*$ is conformally invariant in six dimensions. This leaves the overall volume of the compactification manifold undetermined. These conditions can be derived from a superpotential given by
 the Gukov-Vafa-Witten (GVW) form \cite{Gukov:1999ya}
\begin{equation}
W=\int_{\cal M} \Omega \wedge G_3.
\end{equation}
This superpotential depends on the complex structure moduli through $\Omega$ and is independent of the K\"ahler moduli.

The $N=1$ supergravity scalar potential is given by
\bea
V =e^{K}\left( \sum_{a,b} g^{a\bar b} D_{a}W \overline {D_b W}
- 3|W|^2 \right),
\label{treepot}
\eea
where $a, b$ run over all the moduli. Due to the no-scale structure of the K\"ahler potential (~\ref{kahler1}), sum over K\"ahler moduli cancels the term $3|W|^2$ and the potential (\ref{treepot}) reduces to the no scale structure \cite{noscaleSUGRA}
\bea
V_{\text{no-scale}} = e^{K}\left(\sum_{i,j} g^{i \bar j} D_i W
\overline{D_j W} \right),
\eea
where $i, j$ run over the dilaton and complex structure moduli. Accordingly, the dilaton and complex structure moduli can be stabilized
in a supersymmetric minimum by solving the equation $D_a W=0$ which may have a solution for generic choice of the flux. In this case, $W=W_0$ at the minimum.

%
%

The above discussion shows that the no-scale structure doesn't fix the value of the volume modulus $T$, i.e., the modulus $T$ is a flat direction. The stabilization of $T$ is Of uttermost importance in order for string theory to make contact with realistic models. This issue will be addressed in the upcoming sections.
%
\section{KKLT and its variants}\label{sec:KKLT}
%
 In order to stabilize the volume modulus $T$, a non-perturbative superpotential was considered by Kachru, Kallosh, Linde and Trivedi (KKLT) \cite{Kachru:2003aw}. The source of these non-perturbative terms could be either D3-brane instantons or gaugino condensation from the non-abelian gauge
sector on the D7-branes. As advocated in the previous section, the dilaton and the complex structure moduli are stabilized at a high
scale by the flux induced superpotential and hence their contribution to the superpotential is a constant $W_0$. Thus, the total effective
superpotential is given by
\be
W = W_0 + A e^{-a T}.
\label{adssup}
\ee
The coefficient $a=2\pi$ or $2\pi / N$ is the correction arising from $D3$ instantons
or $SU(N)$ gaugino condensation, and $A$ is constant of order ${\cal O}(1)$.
In addition, the K\"ahler potential is given by
\be
K = - 3 \ln[T + \overline{T}].
\ee
Here, $T=\tau+i\psi$, with $\tau$ is  the volume  modulus of the internal manifold and  $\psi$ is the axionic part. A Supersymmetric minimum is obtained by solving the equation
\bea
\label{susyAdS}
D_T W = 0 
\ \ \ \ \ \Rightarrow \ \ \ \  W_0=- Ae^{-a\tau_{0}}(1+ {2\over 3} a \tau_{0}),
\eea
with $\tau_0$ is the value of $\tau$ that minimizes the scalar potential.

%
\begin{figure}[htbp]
\begin{center}
\includegraphics[width=0.7\linewidth]{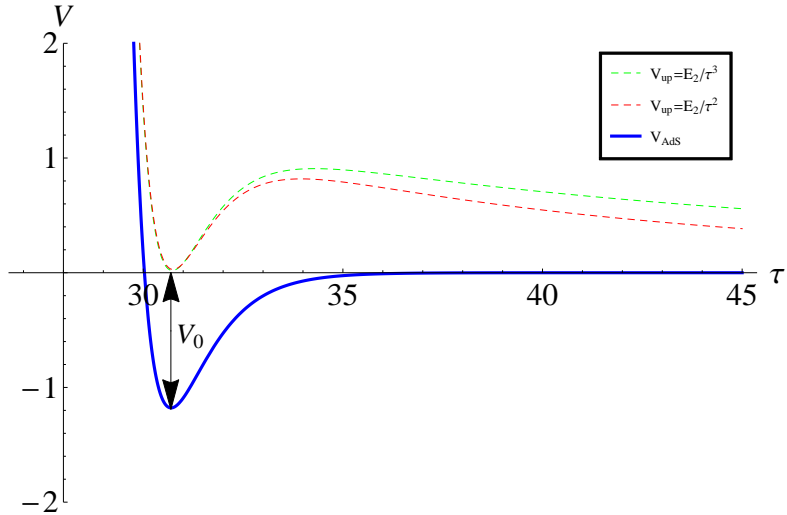}
\end{center}
\vspace*{2mm}
\caption{The scalar potential $V(\tau)$ (multiplied by $10^{29}$) with $W_0 =- 10^{-12}$, $A=1$, $a =1$. The blue curve shows the AdS minimum, while the green and red curves exhibit the
 uplifting to dS minimum via $\delta V = \frac{E_3}{\tau^3} , \frac{E_2}{\tau^2}$, respectively, with $E_3=3.5\times 10^{-25}$ and $E_2=1.13\times 10^{-26}$.}
\label{vacua}
\end{figure}
Substituting this solution in the potential
\be
V= e^{K}\left(  \frac{3}{(T+\overline T)^2} |D_{T}W |^2
- 3|W|^2 \right),
\ee
one finds the following negative  minimum
\be
V_{0}^{AdS}=\big(-3 e^{K}|W|^2\big)|_{\tau_{0}}= -{a^2 A^2 e^{-2a \tau_{0}}\over 6  \tau_{0}}.
\ee
The scalar potential as a function of $\tau=$ Re$(T)$  is given by\footnote{The imaginary part Im$(T)=\psi$ is frozen at zero.}
\be
V(\tau)=\frac{a A e^{-2 a \tau} \left(A (a \tau+3)+3 W_0 e^{a \tau}\right)}{6 \tau^2}.
\ee

It is important to uplift this AdS minimum to a Minkowski or a dS minimum in order to have realistic models.
The uplift of the above AdS vacuum to a dS one will break SUSY where one needs another contribution to the potential which usually has dependence like $\tau^{-n}$ \cite{Linde:2011ja}
\be
\delta V(\tau)\approx ~|V_0^{AdS}|~\frac{\tau_0^n}{\tau^n} .
\ee
In this case a new minimum is obtained due to shifting $\tau_0$ to $\tau_0'=\tau_0+\varepsilon$, where $\varepsilon$ is given by
\be\label{eq:min-shift}
\varepsilon\simeq\frac{1}{a^2 \tau_0} .
\ee
Since the consistency of the KKLT requires that $\tau_0 ,a \tau_0 \gg 1$ \cite{Kachru:2003aw}, the shift in the minima is much small and we can calculate physical
quantities such as masses in terms of $\tau_0$.
%

 There are many proposals for such uplifting, e.g., adding anti-D3-branes \cite{Kachru:2003aw}, D-term uplift \cite{Burgess:2003ic, Villadoro:2005yq, Achucarro:2006zf, Choi:2006bh, Dudas:2006vc, Haack:2006cy, Burgess:2006cb}, F-term uplift \cite{Kallosh:2006dv, Lebedev:2006qq} and K\"ahler uplift \cite{Dudas:2006vc, Westphal:2006tn, Rummel:2011cd, Louis:2012nb}.
In the original KKLT scenario \cite{Kachru:2003aw}, some anti-D3-branes were added which contributes and additional  part to the scalar potential
\be
\delta V = { E_3 \over \tau^3 }.
\label{uplift1}
\ee
%

%
%

One of the drawbacks of this mechanism is that SUSY is broken explicitly due to the addition of the anti-D3-branes.
In this case, the effective 4D theory cannot be recast into the standard form of 4D supergravity and this in turn makes it very difficult to have a low energy effective theory \cite{Burgess:2003ic}.
An uplifting mechanism  via a D-term scalar potential was proposed in \cite{Burgess:2003ic} where the possible fluxes of gauge fields living on the D7 branes were used.
In this case, the fluxes induce a term in the 4D effective action of the form
\be
T_7 \int_\Sigma d^4y \sqrt{g_8} F^{mn} F_{mn} = \frac{2 \pi  E^2}{\tau^3} ,
\ee
with $\Sigma$ is the 4-cycle on which D7 branes wrap, $T_7$
is the tension of D7-branes and $E$ measures the strength of the flux.
Accordingly, the D-term scalar potential is given by
\be \label{eq:VD}
 V_D\ =\ {g^2_{YM} \over 2}  D^2 = \frac{2\pi}{\tau}
 \left(\frac{E}{\tau} +\sum{q_i |C_i|^2}\right)^2  ,
\ee
where $C_i$ are charged matter fields with charges $q_i$.
These fields can be minimized at $C_i = 0$ and accordingly, the full potential will be
\be
 V\ = \ V_F + V_D
\ee
with $V_D=2\pi E^2/\tau^3$ and we end with similar behavior to the KKLT potential.
%

%
\begin{figure}[htbp]
\begin{center}
\includegraphics[width=0.5\linewidth]{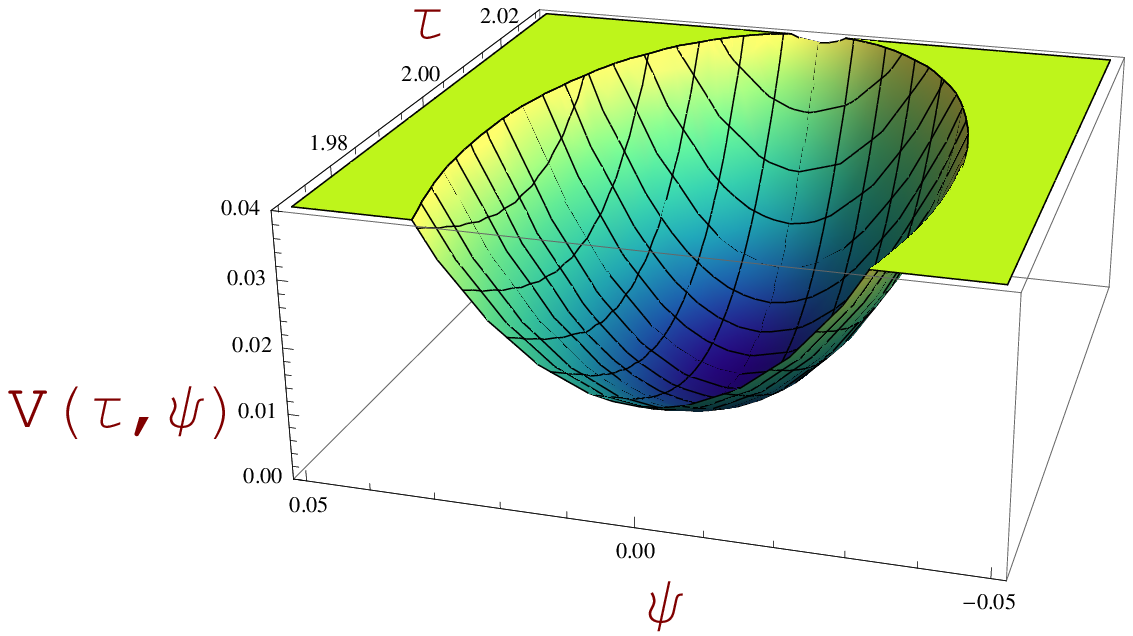}~~\includegraphics[width=0.5\linewidth]{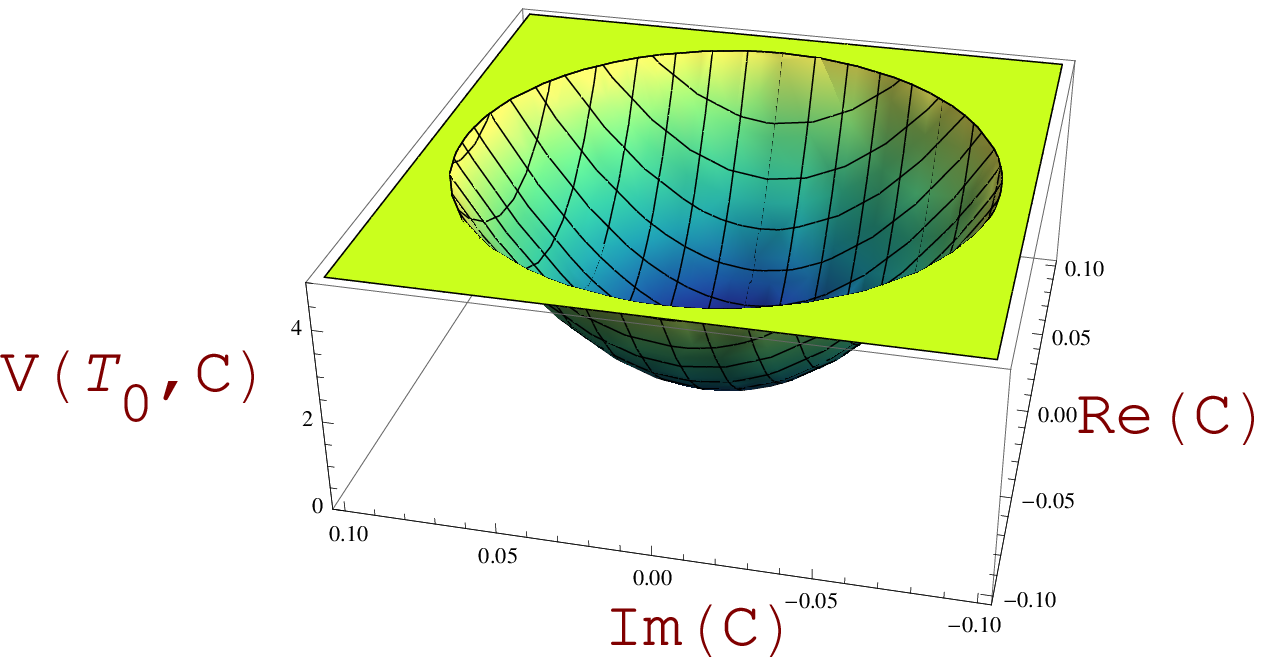}
\end{center}
\vspace*{2mm}
\caption{The minimum of the potential of the scenario \cite{Lebedev:2006qq} at $T=2, C=0$. The left panel corresponds to $V(T,0)$ rescaled by $\frac{1}{\epsilon^2}$ and the right panel corresponds to
corresponds to $V(T_0,C)$ rescaled by $\frac{10^{4}}{\epsilon^2}$.}
\label{fig:dS-matter}
\end{figure}

Another approach for uplifting uses the F-term \cite{Kallosh:2006dv, Lebedev:2006qq}. In this case, SUSY will be broken spontaneously in the F-term moduli sector which in turn generates an uplift term for the AdS KKLT stabilized volume.
For example, in \cite{Lebedev:2006qq} the K\"ahler potential contains a modulus $T$ and a matter field
$C$ and has the form
\be
 K=-3 \ln (T+\overline{T}) + \vert C \vert^2 .
\ee
The effective superpotential \cite{Lebedev:2006qq} is given by
\begin{equation}
 W= \sum_i \omega_i(C) {\rm e}^{-\alpha_i T}  + \phi (C) ,
\label{WLN}
\end{equation}
where  $\omega_i(C)$ and
$\phi (C)$ are functions of the matter fields resulting due to integrating out heavy fields with the index $i$ runs over the gaugino condensates.
The scalar potential will be minimized at \cite{Lebedev:2006qq}
\be
C=0,  \qquad  T=T_0 .
\ee
Figure \ref{fig:dS-matter} shows the shape of the potential near the minimum $T=T_0,C=0$.
This model is different from the approach studied in \cite{Awad:2005gr}, where the effects of charged chiral fields that resides on D3 and D7 branes \cite{Ibanez:1998rf} were considered.
In that case, new $T$-dependence will be generated in the K\"ahler potential
\be
 K~=~-3 \ln (T+\overline{T}) -3 \ln (S+\overline{S})+ \frac{\vert \langle C_3\rangle \vert^2}{(T+\overline{T})} + \frac{\vert \langle C_7\rangle \vert^2}{(S+\overline{S})} \;,
\ee
where $C_3,  C_7$ are charged chiral matter fields. In addition the superpotential does not contain any non-perturbative effects
\be
 W~=~A+BS.
\ee
Therefor, an AdS minimum is obtained which is uplifted by a D-term associated with the gauge symmetry of the matter fields.

\begin{figure}[htbp]
\begin{center}
\includegraphics[width=0.7\linewidth]{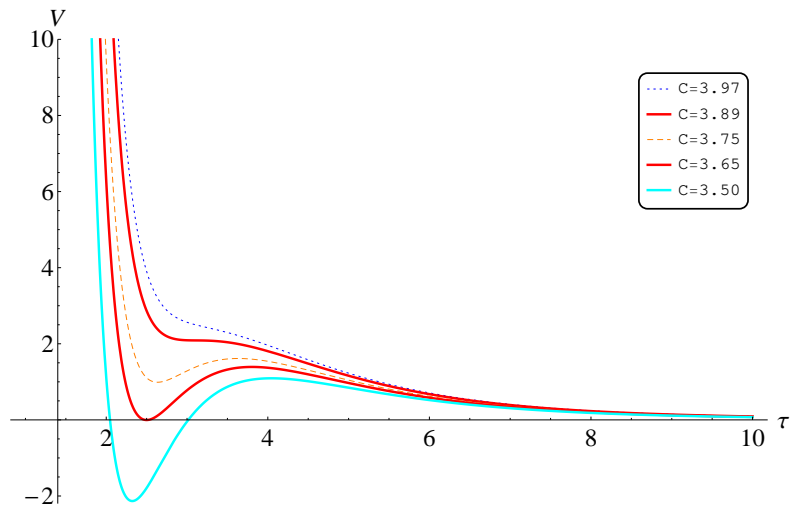}
\end{center}
\vspace*{2mm}
\caption{Potential of one modulus model for the K\"ahler uplift. The red curves correspond to the limits on the parameter $C$ for a dS vacuum, while the dotted one corresponds to $C$
in the destabilization region.}
\label{fig:CondOnC}
\end{figure}

Another way for uplifting to dS vacua of the volume stabilized moduli, is by {\it K\"ahler uplift models} where the perturbative corrections to the K\"ahler potential will paly an important role in constructing dS vacua \cite{Dudas:2006vc, Westphal:2006tn, Rummel:2011cd, Louis:2012nb}.
 Now we consider a model proposed in  \cite{Rummel:2011cd} where $h^{1,1}=1$ and $h^{2,1}>1$ so that the Euler characteristic is $\chi = 2 (h^{1,1} - h^{2,1}) < 0$. In this case, the K\"{a}hler potential and superpotential are
\bea\label{ModelDef}
K &=& - 2 \ln \left( \hat{\mathcal{V}} + \alpha'^3 \frac{\hat\xi}{2} \right) ,\\
W &=& W_{0} + \sum_i A_i e^{- a_i T_i},
\eea
where $\hat{\cal{V}}= \gamma (T + \bar{T})^{3/2}$ is the normalized volume, with $\gamma= \sqrt{3}/(2\sqrt{\kappa})$ and
 \be
 \hat\xi= - \frac{\chi({\cal M})  \zeta(3)}{4 \sqrt{2} (2 \pi)^3}   (S+\bar{S})^{3/2}.
 \ee
Accordingly, the $\tau$ dependance of $N=1$ supergravity scalar potential is given by \cite{Balasubramanian:2004uy, Becker:2002nn, Rummel:2011cd}
\be
V(t) = e^{K} \left( K^{T\bar{T}} \left[ a^2 A^2 e^{- 2 a t} + (- a A e^{-a t} \overline{W K_T} + c.c) \right] + 3 \hat\xi \frac{\hat{\xi}^2+7\hat{\xi}\hat{\mathcal{V}}+\hat{\mathcal{V}}^2}{(\hat{\mathcal{V}}-\hat{\xi})(\hat{\xi}+2\hat{\mathcal{V}})^2} |W|^2 \right),
\label{eq:V-Kup}
\ee
where $\psi$ stabilizes at $\psi = n \pi / a$ for $n=0,1,\ldots~$.
Using the approximation $\hat{\mathcal{V}} \gg \hat{\xi}$ and $|W_0| \gg A e^{- a t}$ and defining the quantities
\be
x=a \cdot t, \qquad  C = \frac{-27 W_0 \hat\xi a^{3/2}}{64 \sqrt{2} \gamma A},
\ee
then the scalar potential will be simplified to the form
\begin{equation}
V(x) \simeq \frac{- W_0 a^3 A}{2 \gamma^2} \left( \frac{2 C}{9 x^{9/2}} - \frac{e^{-x}}{x^2} \right).
\label{VFt2termx}
\end{equation}
In order to have stable dS vacuum, $C$ must satisfy the constraint \cite{Rummel:2011cd}
\be
3.65 \lesssim C\lesssim 3.89 .
\label{eq:dSCond}
\ee
This is clarified in Figure~\ref{fig:CondOnC}, where for values of $C <3.65$, we have AdS minimum and for values of $C > 3.89$, the volume is destabilized.
%
%

%
\section{Moduli stabilization in Large Volume Scenario}\label{sec:LVS}
%
Another alternative scenario for moduli stabilization based on a large volume scenario has been proposed by Quevedo, \etal~ \cite{Balasubramanian:2005zx, Conlon:2005ki}
in order to overcome some of the drawbacks of the KKLT model.
Basically, increasing the number of K\"ahler
moduli will worsen the situation when $\alpha'$ corrections are neglected.
The LVS was built on the proposal that the number of complex structure moduli is  bigger than the number of K\"ahler
moduli, i.e., $h^{2,1}>h^{1,1}>1$ as well as the inclusion of $\alpha'$ corrections.
The ${\cal O}(\alpha'^3)$ contribution to the K\"ahler potential (after integrating out the dilaton and the complex structure moduli) is given by \cite{Becker:2002nn, Grimm:2004uq, Balasubramanian:2004uy}
\bea \label{LVSKahler}
K_{\alpha'} = -2 \log
\left[e^{-3\phi / 2} \mathcal{V} + \frac{\xi}{2}
\left(\frac{\left(S +
\bar{S}\right)}{2}\right)^{3/2}\right] +K_{cs} ,
\ee
with $\xi = -
\frac{\zeta\left(3\right) \chi\left({\cal M}\right)}{2(2 \pi)^3}$ and $\phi$ is the type-IIB dilaton. Here $\mathcal{V}$ is defined as the classical volume of the manifold ${\cal M}$ which is given by
\be
{\cal V} = \int_{\cal M} J^3 = {1\over 6} \kappa_{ijk} t^i t^j t^k,
\ee
where $J$ is the K\"ahler class and $t_i$ are moduli that measure the areas of 2-cycles with $i=1,2,\ldots,h_{1,1}$.
The complexified K\"ahler moduli are defined as $T_j \equiv \tau_j + i \psi_j$ with $\tau_j $ are the four-cycle moduli defined by the relation
\be
\tau_j = \partial_{{t}_j} {\cal V} = {1\over 2} \kappa_{jkl} {t}^k {t}^l~.
\label{4to2cycles}
\ee

Since the superpotential is not renormalized at any order in perturbation theory, it will not receive  $\alpha'$ corrections.
But there is a possibility of non-perturbative corrections, which may depend on
 the K\"ahler moduli (as in the KKLT model) via D3-brane instantons or gaugino condensation from wrapped D7-branes. Accordingly, the superpotential
  is given by
\be
\label{npsuperpot}
W = W_0 + \sum_i A_i e^{- a_i T_i},
\ee
where $A_i$ is a model dependent constant and
again $W_0$ is the value of the superpotential due to the geometric flux after stabilizing the dilaton and the complex structure moduli.
In this respect, the scalar potential will take the form \cite{Becker:2002nn, Balasubramanian:2004uy}
\bea
\label{potential}
V & = & e^K \left[ K^{T_j \bar{T}_k} \left(a_j A_j a_k \bar{A}_k e^{ \left(a_j T_j + a_k \bar{T}_k\right)}
+   a_j A_j e^{i a_j T_j} \bar{W} \partial_{\bar{T}_k} K - a_k \bar{A}_k e^{-i a_k \bar{T}_k}
W \partial_{T_j} K \right) \right. \nonumber \\
& + & \left. 3 \xi \frac{\left(\xi^2 + 7\xi \mc{V} +
\mc{V}^2\right)}{\left(\mc{V} - \xi\right)\left(2\mc{V} +
\xi\right)^2}
|W|^2\right] \\
& \equiv & V_{np1} + V_{np2} + V_{\alpha'}. \nonumber
\eea
where the $\alpha'$ correction is encoded in $V_{\alpha'}$.
The simplest example that can realize the notion of LVS \cite{Balasubramanian:2005zx, Conlon:2005ki} is the orientifold of $P^4_{[1,1,1,6,9]}$ for which $h^{1,1}=2$ and $h^{2,1} = 272$ and therefor the volume is given by
\be
{\cal V} = \frac{1}{9\sqrt{2}}\left(\tau_5^{\frac{3}{2}} -
\tau_4^{\frac{3}{2}}\right),
\ee
where the volume moduli are $T_4=\tau_4+i\psi_4$ and $T_4=\tau_5+i\psi_5$ and the link to $t_i$ is given by $\tau_4=\frac{t_1^2}{2}$ and $\tau_5=\frac{(t_1+6 t_5)^2}{2}$.
 In this respect, the K\"ahler potential and the superpotential, after fixing the dilaton and complex structure moduli, are given in the string frame by
\be
\begin{split}
K &= K_{cs}-2\log\left(\mc{V}+\frac{\xi}{2}\right),\\
W& = W_0 + A_4 e^{- \frac{a_4}{g_s} T_4} + A_5 e^{- \frac{a_5}{g_s} T_5}.
\end{split}
\ee
\begin{figure}[htbp]
\begin{center}
\includegraphics[width=0.6\linewidth]{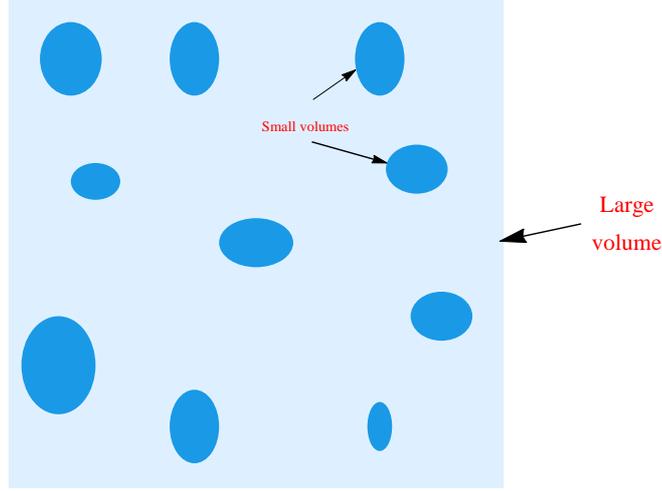}
\end{center}
\vspace*{2mm}
\caption{ Swiss Cheese structure in the Large Volume Scenario.}
\label{swiss-cheese}
\end{figure}

In the large volume limit,   ${\cal V} \sim  \tau_5 \gg\tau_4 > 1$, the behavior of the scalar potential is given by \cite{Balasubramanian:2005zx, Conlon:2005ki}
\be
\label{Vgen}
V(\mc{V}, \tau_4)= \frac{ \sqrt{\tau_4} (a_4 A_4)^2 e^{-2 a_4 \tau_4/g_s}}{\mc{V}}
- \frac{ W_0 \tau_4a_4 A_4 e^{-a_4 \tau_4/g_s}}{\mc{V}^2}  + \frac{\xi W_0^2}{\mc{V}^3}.
\ee
 Minimizing the potential (\ref{Vgen})
\be
\frac{\partial V}{\partial \mc{V}} = \frac{\partial V}{\partial \tau_4} = 0 .
\ee
and solving the two equations in the two variables, $\tau_4, \mc{V}$ one can get
one equation in $\tau_4$
\be
\label{tausoln}
\left(1 \pm \sqrt{1 - \frac{3 B_3 B_1}{B_2 \tau_4^{\frac{3}{2}}}} \right)\left( \frac{1}{2} - 2 a_4 \tau_4 \right)
= (1 - a_4 \tau_4),
\ee
where
\be
B_1 \sim a_4^2 \vert A_4 \vert ^2, \qquad B_2 \sim a_4 \vert A_4
W_0 \vert, \qquad
B_3 \sim \xi \vert W_0 \vert^2.
\ee
Using the assumption $a_4\tau_4\gg 1$, which is necessary to neglect higher order instanton corrections \cite{Balasubramanian:2005zx}, the solution is given by
\bea
\tau_4 & = & \left( \frac{4 B_3 B_1}{B_2^2} \right)^{\frac{2}{3}},
\nonumber \\
\mc{V} & = & \frac{B_2}{2 B_1} \left( \frac{4 B_3 B_1}{B_2^2}
\right)^{\frac{1}{3}}
e^{a_4 \left( \frac{4 B_3 B_1}{B_2^2} \right)^{\frac{2}{3}}}.
\eea
Substituting for $B_i$ by their expressions we have
\be
\label{TauVolFinal}
\tau_4 \sim (4 \xi)^{\frac{2}{3}}, \qquad
\mc{V} \sim \frac{\xi^{\frac{1}{3}} \vert W_0 \vert}{a_4 A_4} e^{a_4 \tau_4/g_s}.
\ee
Therefor potential possesses an AdS minimum at exponentially large volume $\mc{V}$ since it approaches zero from below in the limit $\tau_5\to \infty$ and $\tau_4\propto \log(\mc{V})$.
Namely, in the latter limit, the potential have the form
\be
V=\frac{W_0^2}{{\cal V}^3}\left(C_1\sqrt{\ln(\cal V)}-C_2 \ln(\cal V) + \xi \cal V\right).
\ee
Therefor, the negativity of the potential require a very large $\ln(\cal V)$.
Still one has to uplift this minimum  by one of the mechanisms mentioned in Section (\ref{sec:KKLT}).
This result can be generalized to more than two moduli where one of them  takes a large limit while the other moduli stays small.
This structure will form what is called the {\it Swiss-cheese} form of the CY manifold as depicted in Figure~\ref{swiss-cheese}. In this respect the volume will take the form
\be
\mc{V} = \tau_b^{3/2}-\sum_i\tau_{s,i}^{3/2}.
\ee
%

%
\section{SUSY breaking and phenomenological consequences}\label{sec:SUSY-break}
%
In this section we will study SUSY breaking in the moduli sector and the properties of the corresponding soft terms\footnote{More general expressions of soft term for generic superpotentials and K\"ahler potentials in supergravity and string models, are given in  \cite{Brignole:1997dp}.} that are induced in the observable sector then we summarize the phenomenological consequences.
SUSY breaking in models of KKLT compactification type with phenomenological consequences has been extensively studied in \cite{Choi:2006bh, Linde:2011ja, Choi:2005ge, Endo:2005uy}, while for LVS type, SUSY breaking was studied in \cite{Conlon:2005ki, Allanach:2005pv, Conlon:2006us, Conlon:2006wz, Angus:2012dd, Aparicio:2014wxa}.

{\bf i) Soft terms in KKLT scenario:}\\

As shown in models of KKLT type in Section (\ref{sec:KKLT}), SUSY is broken by one of the uplifting mechanisms. The gravitino mass at the dS minimum is given by
\bea
m_{3/2} =e^{K/2} |W|\Big|_{dS}  \Rightarrow      m_{3/2} \simeq  \frac{a~A}{3(2\tau_0)^{1/2}}e^{-a\tau_0}  \simeq  \frac{W_0}{(2\tau_0)^{3/2}}.
\eea
Therefor we have gravitino mass of order TeV if $ (a\tau_0)\sim 32$ and hence $W_0\sim 10^{-12}$.
In terms of the shifts $\varepsilon$ (\ref{eq:min-shift}) of $\tau_0$, an approximate expression for $D_T W$ near the dS minimum is given by
\be
D_T W(\tau_0+\varepsilon) = (D_T W)_\tau \varepsilon \simeq W_{T,T}~\varepsilon= \frac{3\sqrt{2}}{a\sqrt{\tau_0}}~m_{3/2},
\ee
which is the same order as the gravitino mass.
Accordingly, the soft SUSY breaking terms are given by
\bea
m_0^2&=& \frac{|W|^2}{(2\tau)^3} \Big|_{dS} =m^2_{3/2},\n
m_{1/2} &=& \frac{\sqrt{2\tau}}{6} D_T W(T)~ \frac{\partial}{\partial T}\ln ({\rm Re} f^*)\Big|_{dS} \simeq  \frac{m_{3/2}}{a\tau},\n
A_0 &=& - \frac{1}{\sqrt{2\tau}} {\bar D}_{\bar T}  \overline{W_h} \Big|_{dS}  =  - \frac{3 m_{3/2}}{a\tau},
\eea
where the gauge kinetic function $f_{ab}$ can be chosen such that $f_{ab}(T) =f(T) \delta_{a b}$, which will lead to universal gaugino masses, is considered to have a linear dependance on the modulus field that can be derived from the reduction of Dirac-Born-Infeld action for an unmagnetized
brane \cite{Conlon:2006wz}.
%
%
%
%

The soft terms indicate that SUSY breaking in KKLT is a special example of the constrained MSSM (CMSSM), where all the soft terms are given in terms of one free parameter ($m_{3/2}$) which is of order TeV. Note that $a \tau_0$ is fixed as $a \tau_0\simeq 32$. It is well known that this type of soft terms can't account for the experimental constraints imposed by the Large Hadron Collider (LHC) and relic abundance of the lightest SUSY particle. Even if we relax the Dark matter constraints, the mass limit ($\sim 125$ GeV) and gluino mass limit ($\gtrsim 1.4$ TeV)
will imply $m_{3/2}\simeq {\cal O} (30)$ TeV. Thus all SUSY spectrum will be quite heavy which is beyond the LHC sensitivity. A feature of this model is the fact that gauginos are lighter than the sfermions by at least one order of magnitude. However if one checks the parameter space for such set of soft terms, he finds that tadpole equations at the TeV scale are not satisfied. Namely, the condition \cite{Chakraborti:2014fha}
\be\label{eq:tadepol}
\mu^2+\frac{M_Z^2}{2}\simeq -0.1 m_0^2+2m^2_{1/2},
\ee
can not account for positive $\mu^2$.
Therefor the above set of soft terms failed to describe TeV scale phenomenology.
%
%
%
\\

{\bf ii) Soft terms in LVS scenario}\\

\begin{figure}[htbp]
\begin{center}
\includegraphics[width=0.5\linewidth]{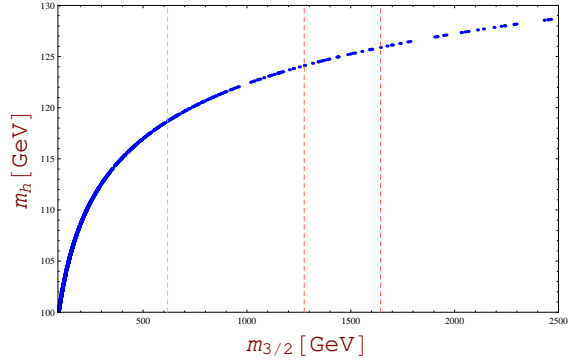}
\end{center}
\vspace*{2mm}
\caption{Higgs mass $m_h$ as a function of the gravitino mass $m_{3/2}$. The region in left to the green line is disallowed by the gluino mass constraint, while the area between the red lines shows the region for which the Higgs mass lies between $124-126$ GeV.}
\label{fig:LVS-scan}
\end{figure}

Let us now consider the same scenario for SUSY breaking in LVS model, namely with $P^4_{[1,1,1,6,9]}$ geometry. In this case we have
\be
\begin{split}
K &= -2\log\left(\mc{V}+\frac{\xi}{2}\right),\\
W &= W_0 + A_4 e^{- \frac{a_4}{g_s} T_4} + A_5 e^{- \frac{a_5}{g_s} T_5}.
\end{split}
\ee
Therefor the gravitino mass will be given by
\be
m_{3/2} = e^{K/2} \vert {W} \vert \Big|_{dS} \sim \frac{g_s^2
   \vert W_0 \vert}{{\cal V} \sqrt{4 \pi}}  M_p.
\ee
It is remarkable that depending on the large volume $\cal V$, the gravitino mass could be TeV or much larger.
Considering the K\"ahler metric of the observable sector to have the form $\tilde{K}_{\alpha\bar{\beta}}=\tilde{K}_{\alpha} \delta_{\alpha\bar{\beta}}$, with $\tilde{K}_{\alpha}$
 is constant, we have scalar soft masses of the following form
\bea
m_\alpha^2=m_{3/2}^2,
\eea
where we have neglected the very tiny cosmological constant value $V_0$.
In the large volume limit,
${\cal{V}} \sim\tau_5 \equiv \tau_b  \gg \tau_4 \equiv \tau_s\gg 1$, the K\"ahler metric and its inverse are given by
\bea
K_{m\bar{n}}\simeq\left[
  \begin{array}{cc}
   {1\over \cal{V}} ~~&~~ -{1\over {\cal{V}}^{5/3}} \\
    -{1\over {\cal{V}}^{5/3}} ~~&~~ {1\over {\cal{V}}^{4/3}} \\
  \end{array}
\right],\hspace{1.5cm}
K^{m\bar{n}}\simeq\left[
  \begin{array}{cc}
   {\cal{V}} ~~&~~ {\cal{V}}^{2/3} \\
    {\cal{V}}^{2/3} ~~&~~ {\cal{V}}^{4/3} \\
  \end{array}
\right].
\eea
The F-terms can be calculated from
\be
F^m=e^{K/2}K^{m\bar{n}} D_{\bar{n}}W.
\ee
 Hence the approximate dependence of the F-terms on the volume, is given by
\bea
F^4  \sim {1\over{\cal{V}}}
,\hspace{1.5cm}
F^5 \sim {1\over{\cal{V}}^{1/3}}.
\eea
Consider a linear dependence of the gauge kinetic function on the K\"ahler moduli \cite{Conlon:2006wz}, therefor the gaugino masses are given by
\bea
m_{1/2}\sim \frac{g_s^2 W_0 M_p}{\sqrt{4 \pi}} \left(\frac{1}{\cal V}+\frac{1}{{\cal V}^{5/3}} \right).
\eea
Also the universal A-term is given by
\bea
A_0\simeq F^{m}K_m\sim \frac{-g_s^2 W_0 M_p}{\sqrt{4 \pi}{\cal V}}.
\eea
Accordingly, at large volumes ${\cal V}\sim 10^{13}-10^{15}$ the soft masses are related to the gravitino mass as follows
\bea
m_0 &\simeq& m_{1/2}= m_{3/2},\n
A_0&\simeq& - m_{3/2} .
\eea
%
These soft terms are generated at string scale, therefor we have to run them to the electroweak (EW) scale and impose the EW symmetry breaking conditions to analyze the corresponding spectrum.
In this case the condition (\ref{eq:tadepol}) reads
\be\label{eq:tadepol-lvs}
\mu^2+\frac{M_Z^2}{2}\simeq -0.1 m_{3/2}^2+ 2 m_{3/2}^2,
\ee
which can be satisfied easily at the TeV scale. The above set of soft terms is special case of CMSSM. Therefor, the Higgs mass limit requires $m_{1/2}$ to be of order 1.5 TeV, as shown in  Figure~\ref{fig:LVS-scan}, where the Higgs mass $m_h$ is plotted versus the gravitino mass $m_{3/2}$.
With such heavy values of $m_0$ and $m_{1/2}$, the SUSY spectrum will be quite heavy and the relic abundance of the lightest neutralino is not consistent with Planck's results  \cite{Abdallah:2015hza}.

\section{Cosmological consequences}\label{sec:mod-cosmo}
%
One of the important consequences of the moduli stabilization is its effect on inflation. It turned out that the scalar potential of the modulus field in KKLT is not suitable to account for single-field inflation since the modulus potential is not flat enough to allow for slow roll \cite{BlancoPillado:2004ns}. Namely, the real part is not protected by a shift symmetry which will result in an $\eta$-problem. Although the axionic part is subjected to shift symmetry, the field dependent mass matrix is not diagonal with a significant mixing. Therefor the shift symmetry is violated and the $\eta$-problem persists.\footnote{For a review for string inflation in the light of recent observations see for example  \cite{Burgess:2013sla}.}

In this section we discuss how this problem can be evaded in what is called the {\it racetrack superpotential}. Also, inflation via k\"ahler moduli is discussed.
In addition we explain the destabilization problem that arises in KKLT and LVS. Finally we study the effect of moduli backreaction on inflationary scenarios.
%
\subsection{ Racetrack inflationary model}\label{ssec:racetrack-inf}
%
In the racetrack model, the superpotential is given by \cite{BlancoPillado:2004ns}
\be
W=W_0+Ae^{-a T}+B e^{-b T}.
\ee
where the additional non-perturbative term may be obtained from gaugino condensation in a theory with a product of gauge groups such as $SU(N)\times SU(M)$ gauge group, hence $a=2 \pi/M$  and $ b=2\pi /N $. This is known as the racetrack model. In this case, the global SUSY minimum is located at
\be
T_0=\frac{M N}{M-N}\log\left(- \frac{M B}{NA} \right)
\ee
\begin{figure}[htbp]
\begin{center}
\includegraphics[width=0.5\linewidth]{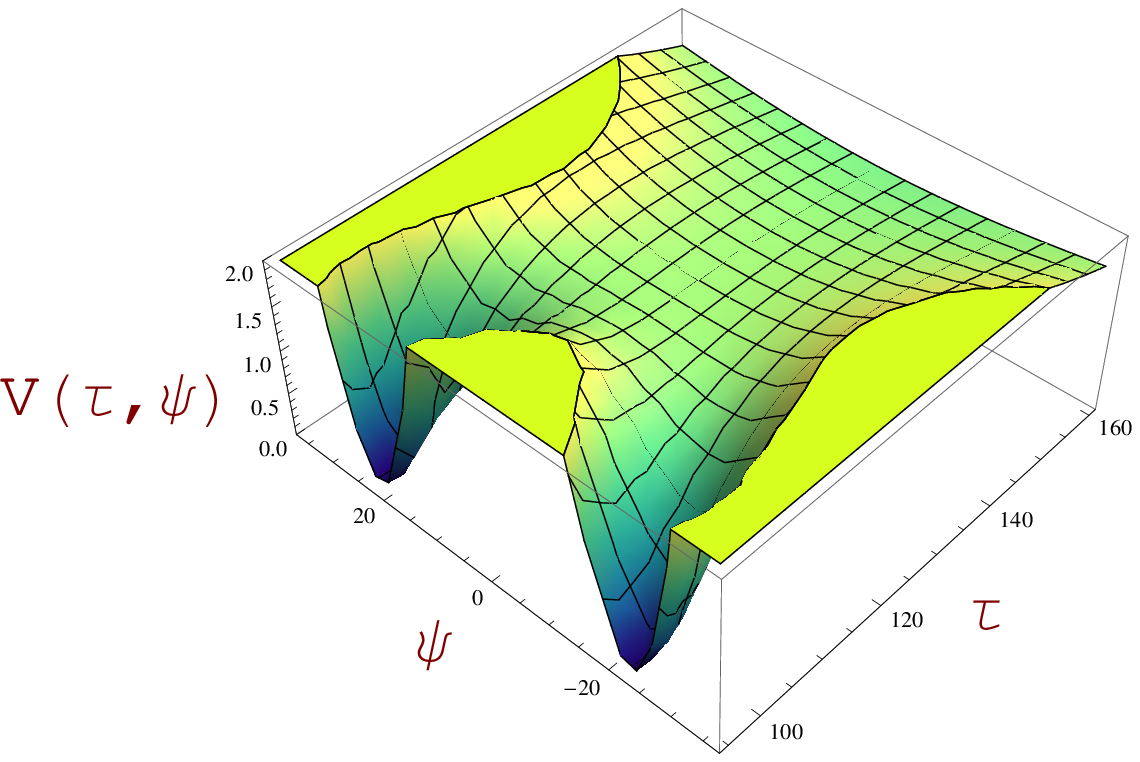}~\includegraphics[width=0.5\linewidth]{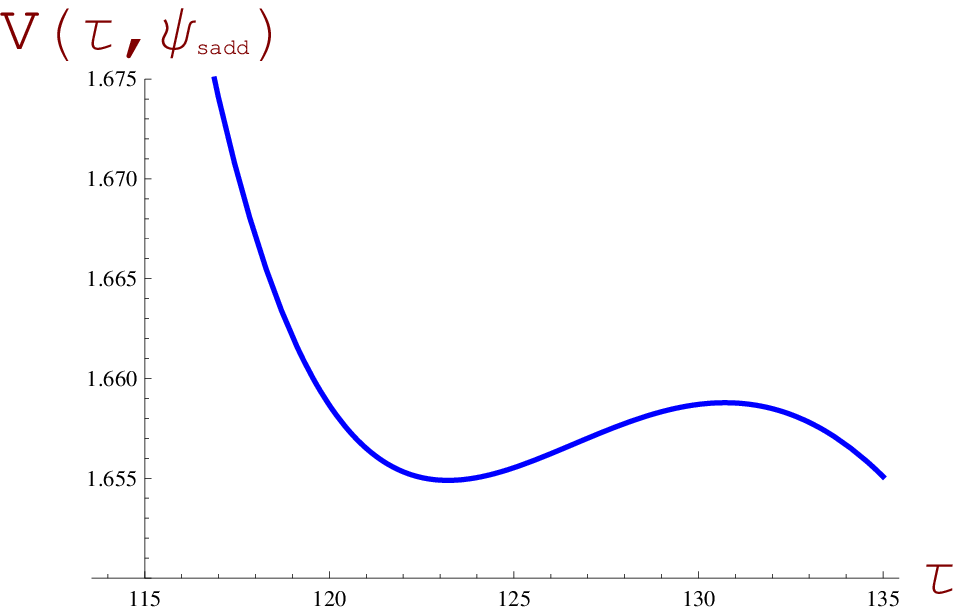}
\end{center}
\vspace*{2mm}
\caption{The racetrack scalar potential (left panel) and (right panel) corresponds to the variation of the potential in the neighborhood of the saddle point ( all are multiplied by $10^{16}$).The parameter values are $A=0.02$, $B=0.035$, $a=\frac{2\pi}{100}$, $b=\frac{2\pi}{90}$, $E_2=4.14668 \times 10^{-12}$ and
$W_0 = -4 \times 10^{-5}$.}
\label{fig:racetrack-pot}
\end{figure}

The inflationary potential is given by the F-term scalar potential added to the uplifting term as follows
\be
\begin{split}
V_{\text{inf}}&=\frac{E_2}{\tau^2}+ \frac{e^{-4 \tau (a+b)}}{6 \tau^2} \Bigg\{a A^2 (a \tau + 3 ) e^{2 \tau (a+2 b)}\\[3pt]
&\quad + e^{3 \tau (a+b)} \Big[A B (2 a b \tau+3 (a+b)) \left(\cos  (\tau  (a-b))\right)+3 a A W_0 e^{b \tau} \cos  (a \psi )\\[3pt]
&\quad +3 b B W_0 e^{a \tau} \cos  (b \psi)\Big]+
 b B^2 (b \tau+3) e^{2 \tau (2 a+b)}\Bigg\},
\end{split}
\ee
%

Figure \ref{fig:racetrack-pot} shows the scalar potential with two degenerate minima and one saddle point located at $\psi_{\text saddle}=0$. At this saddle point, the potential has minimum in $\tau$ direction and maximum in $\psi$ direction .
The inflaton is considered to be slowly rolling from initial conditions near the saddle point on the inflationary trajectory. Namely, the initial motion is in the $\psi$ direction and the inflationary path is determined numerically \cite{BlancoPillado:2004ns}.

The racetrack inflation model predicts spectral index $n_s=0.96$, inflation scale $M_{\text{inf}}\sim 10^{14}$ GeV and tiny tensor to scalar ratio $r\simeq (\frac{M_{\text{inf}}}{M_{\text{GUT}}})^4\sim 10^{-8}$.

%
\subsection{K\"ahler moduli inflation in the LVS}\label{ssec:Kmod-inf}
%
%
%
The idea of K\"ahler moduli inflation \cite{Conlon:2005jm} is to produce an inflationary potential similar to the form
\bea
V_{\text{inf}}(\varphi)= V_0 (1-k \beta e^{-k \varphi}),
\eea
where $\beta$ and $k$ are positive constants.

In the case of multi-modulus Calabi-Yau geometries, the Calabi-Yau volume can take the form
\bea\label{vol} \mc{V} =\frac{\alpha}{2 \sqrt{2}} \left[ (T + \bar{T})^{3/2} -
\sum_{j=1}^n \lambda_j (T_j + \bar{T}_j)^{3/2} \right],
\eea
where $T=\tau+i\psi$ is responsible for the large volume and $T_j=\tau_j+i\psi_j$ with $\tau_j$ are the blow-ups. Here $\alpha$ and $\lambda_j$ are model dependent positive constants. In this case the K\"ahler potential is given by
\be \label{eq:Kmod-inf-pot}
\mc{K} = \mc{K}_{cs} - 2 \ln \left[ \mc{V} + \frac{\xi}{2}
\right],
\ee
while the superpotential is given by
\be
\label{eq:Kmod-sp}
W = W_0 + \sum_{i=2}^n A_i
e^{-a_i T_i},
\ee
where $a_i = \frac{2 \pi}{g_s N}$. The minimum of the scalar potential exists at the large volume limit $\mc{V}\sim \tau\gg \tau_j$ and the scalar potential has the form \cite{Conlon:2005jm}
\be \label{eq:Kmod-pt}
 V =
\sum_i \frac{8 (a_i A_i)^2 \sqrt{\tau_i}}{3 \mc{V} \lambda_i
\alpha} e^{-2 a_i \tau_i} - \sum_i 4 \frac{a_i A_i}{\mc{V}^2} W_0
\tau_i e^{-a_i \tau_i} + \frac{3 \xi W_0^2}{4 \mc{V}^3}. \ee

The inflation can occur away from the minimum and the inflaton is considered to be one of the small moduli $\tau_n$.
The volume $\mc V$ and the small moduli, other than the inflaton, is guaranteed to stabilize to their minima during the inflation \cite{Conlon:2005jm}, therefor the inflationary potential is given by \cite{Conlon:2005jm}
\bea
V_{\text{inf}}= V_0 - \frac{4 a_n A_n W_0\tau_n e^{-k \varphi}}{{\cal V}^2}  ,
\eea
with $V_0=\frac{\beta W_0}{{\cal V}^3}$ which is constant during the inflation. This model predicts $0,967 > n_s > 0.960$ and $r\sim 10^{-10}$ for number of e-folding $50 < N_e < 60$
with inflationary scale $M_{\text{inf}}\sim 10^{13}$ GeV.
%

%
\subsection{Destabilization problem}\label{ssec:destab}
%
In the KKLT scenario, the modulus mass is given by \cite{Linde:2011ja}
\be
m_\tau^2=\frac{V''(\tau)}{2K_{T\bar{T}}}\big|_{\tau_0}=\frac{2}{9} (D_T W)_\tau (\tau W_{TT}-2 W_T)\big|_{\tau_0}.
\ee
Therefor it is linked to $m_{3/2}$ by the relation
\be\label{KKLT:mod-mass}
m_\tau= \frac{\sqrt{2\tau}}{9} W_{TT} \big|_{\tau_0}= 2a\tau_0m_{3/2} ,
\ee
which is approximately two order of magnitude greater than the TeV scale gravitino mass. In case there is an inflaton field different from modulus field with mass $\sim 10^{13}$ GeV, then the modulus may perturb GUT scale inflation. 
It turns out that a problem arises due to the conflict between the requirement of a high energy scale (GUT scale) inflationary phenomena and low energy (TeV scale) SUSY phenomenology
in the context of KKLT. The conflict is stemming from the constraint \cite{Kallosh:2004yh}
\begin{figure}[htbp]
\begin{center}
\includegraphics[width=0.7\linewidth]{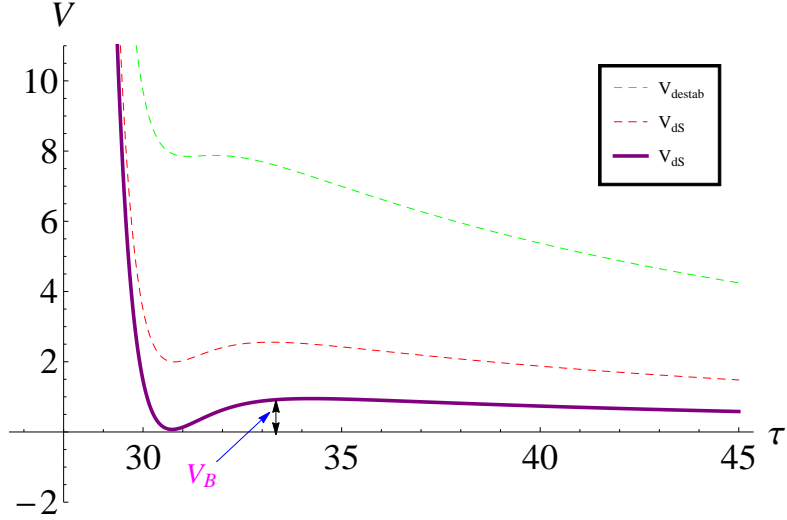}
\end{center}
\vspace*{2mm}
\caption{ The purple curve shows the dS vacuum, while the green one displays the destabilization of the modulus due to large contribution of vacuum energy density.}
\label{fig:destab}
\end{figure}
\be\label{KKLT-H-m32}
m_{3/2}\gtrsim H,
\ee
where $H$ is the Hubble parameter. This constraint originates from the fact that the barrier $V_B$ from the runaway direction, as exhibited in Figure \ref{fig:destab}, is a bit less than the magnitude of the AdS minimum $|V_0^{AdS}|$ since the uplift effect is negligible for large $\tau$. Hence we have
\bea
V_B \lesssim V_0^{AdS}\simeq 3 m_{3/2}^2.
\eea
Therefor, recovering the Planck mass, we have
\be
V_B \sim   m_{3/2}^2 M_p^2.
\ee
If we included an inflaton, there will be a contribution to the overall inflationary potential by a term of the form $V(\varphi)/ \tau^n$ due to the fact that the modulus
couples to all sources of energy. Accordingly, the total inflationary potential at the KKLT dS minimum is given by
\be
V_{\text inf}= V_{dS}(\tau)+ \frac{V(\varphi)}{ \tau^n} \simeq \frac{V(\varphi)}{ \tau^n}.
\ee
In this case there will be a competition between the runaway dependence $\tau^{-n}$ and the barrier $V_B$.
Therefor, protecting the volume modulus from destabilization will impose the condition
\be
V_{\text inf} \lesssim 3 V_B \sim 3 m_{3/2}^2 ~~~\Rightarrow~~~~ m_{3/2}\gtrsim H.
\ee

On the other hand, in the LVS case there is an extra suppression by large volume ${\cal V}$ \cite{Conlon:2008cj}
\be
V_B \sim  \frac{ m_{3/2}^2 M_p^2 }{\cal V} \sim  m_{3/2}^3 M_p,
\ee
and correspondingly we have the constraint
\be
H \lesssim m_{3/2}^{3/2},
\ee
which will not improve the situation much.
In this regard, TeV scale gravitino mass which is favored by low energy phenomenology imply disfavored non-traditional low scale inflation.
The property of $H \lsim m_{3/2}$ seems to be a common property in inflationary models of stabilization scenarios in string theories \cite{Conlon:2008cj, Silverstein:2007ac, Silverstein:2008sg}.
This problem is sometimes called the  {\it Kallosh-Linde} problem. The latter originates from the fact that compatible large scale models of inflation requires a very large gravitino mass where
\be
m_\varphi \ll  H  \lesssim  m_{3/2}.
\ee
%

%
\paragraph{The KL model and strong moduli stabilization:}
%
The problem of destabilization can be evaded in Models of strong moduli stabilization like the KL model \cite{Kallosh:2004yh}.
In this scenario the volume is still determined by one  modulus $T$ but there is additional non-perturbative term contributing to the racetrack
superpotential
\be
 W\ = \ W_0 + A e^{-a T}-Be^{-bT} .
\ee
\begin{figure}[htbp]
\begin{center}
\includegraphics[width=0.7\linewidth]{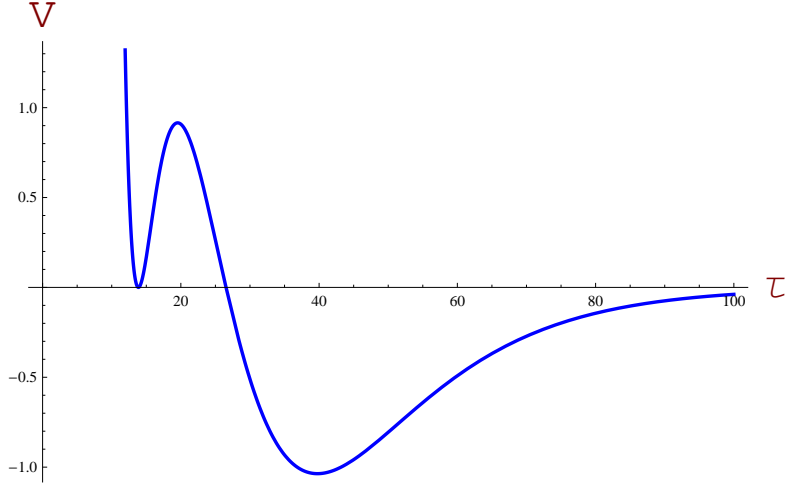}
\end{center}
\vspace*{2mm}
\caption{The F-term scalar potential (multiplied by $10^{7}$) for KL model possesses both metastable Minkowski vacuum which is supersymmetric and another AdS minimum, For $A=B=1$, $a=0.1$, $b=0.05$.}
\label{vkl}
\end{figure}

In this respect, the imaginary part stabilizes at the origin again, whereas the potential of the real part $\tau$ is given by
\bea
V(\tau)=\frac{e^{-2 \tau (a+b)} \left(a A e^{b \tau}-b B e^{a \tau}\right) \left(e^{b \tau} \left(A (a \tau+3)+3 \text{w0} e^{a \tau}\right)-B e^{a \tau} (b \tau+3)\right)}{6 \tau^2}.
\eea
This potential possesses two minima, one is a metastable supersymmetric Minkowski vacuum at
\be
\tau_0 = \frac{1}{a-b}\ln \left( \frac{a A}{b B}\right),
\ee
and the other is a deeper AdS one as shown in Figure~\ref{vkl}. The KL modulus squared mass is given by \cite{Linde:2011ja}
\be\label{KL:mod-mass}
m_\tau^2 = \frac{2a A b B(a-b)}{9} \left( \frac{a A}{b B}\right)^{-\frac{a +b}{a-b}}\ln \left( \frac{a A}{b B}\right).
\ee
Therefor, for $A=B=1$, $a=0.1$, $b=0.05$, we find $m_\tau\sim 4\times 10^{15}$ GeV which means that the volume modulus mass is much larger than inflaton mass ($m_\varphi\sim 10^{13}$ GeV)
 and accordingly will be frozen quickly during the inflation without perturbing the inflaton dynamics. Clearly this is much featured than the KKLT cases since the gravitino mass vanishes at the SUSY minimum. Therefor the hight of the barrier from the runaway direction are independent on the gravitino mass hence the Hubble parameter
is also independent of the gravitino mass. In this case, KL model can account for high scale inflation.

In order to obtain interesting phenomenology at the TeV scale for SUSY breaking without conflict with the high scale inflation requirements,
 a constant shift $\Delta W$ is added to the superpotential \cite{Linde:2011ja} and it is supposed to be of the order of the weak scale.
 In this case the value of $\tau$ minimizing the potential will be shifted to $\tau_0+\delta\tau$. The SUSY vacuum is obtained by solving the equation of motion
\be
D_T W(\tau_0+\delta\tau) =0 ,
\ee
which implies
\be
\delta\tau = \frac{3\Delta W}{2\tau_0 W_{T,T}(\tau_0)}.
\ee
Correspondingly, the AdS minimum is independent of the sign of $\Delta W$ and is approximated to
\be
V_0^{AdS} \simeq -\frac{3(\Delta W)^2}{8\tau_0^3} ,
\ee
where the above expressions are obtained using $W_T(\tau_0)=0$, $W(\tau_0)=\Delta W$ and $W_T(\tau_0+\delta\tau)=\delta\tau ~W_{T,T}(\tau_0)$.
The AdS vacuum can be uplifted to dS one by mechanisms similar to those in KKLT model and hence SUSY will be broken with the gravitino becomes massive.
In this respect, we have the gravitino mass is given in terms $\Delta W$ as
\bea
m_{3/2}\simeq \sqrt{\frac{|V_0^{AdS}|}{3}} = \frac{1}{2\sqrt{2}}\left( \frac{a-b}{\ln \left( \frac{aA}{bB}\right)}\right)^{3/2} |\Delta W|.
\eea
It is worth mentioning that the uplifting effect here is so small and can't exceed the barrier between the dS and the AdS vacua which is one of the strengths of this model.
On the other hand, a weakness of the approach is the lack of interpretation of the origin of the scale of the
shift $\Delta W$ which should be around $10^{-13}$ if we are seeking TeV scale gravitino mass.

\subsection{Impact of string moduli backreaction}\label{ssec:mod-backreaction}
Here we give a brief overview of the effect of string moduli backadation on the inflation models which is linked to SUSY breaking scale. To study such effects on the inflation, models of
strong moduli stabilization are favored like the KL model since the moduli are heavy, in particular the modulus mass is larger than the Hubble parameter.
The moduli backreaction effect on the inflation and its link to SUSY breaking was studied in many research papers \cite{Buchmuller:2014vda, Buchmuller:2014pla, Hebecker:2014kva, Buchmuller:2015oma, Dudas:2015lga}. For a single modulus field case, the impact of the stabilized volume modulus field on large and small field inflation was studied in \cite{Buchmuller:2014vda}. In this respect, the total K\"ahler potential and superpotential are given by
\bea
K &=&K_{\text{mod}}(T+\bar{T})+ K_{\text{inf}}(\phi_\alpha,\bar{\phi}_{\bar{\alpha}}) =-\kappa \log(T+\bar{T})+ K_{\text{inf}},\\
W &=& W_{\text{mod}}(T)+ W_{\text{inf}}(\phi_\alpha),
\eea
where $\phi_\alpha$ are chiral superfields related to the inflationary scenario and the constant $\kappa=1$ for heterotic dilaton and $\kappa=3$ for type IIB K\"ahler modulus. Accordingly the F-term scalar potential can be written as
\be
V = e^{K_{\text{inf}}} V_{\text{mod}}(T)+ (2\tau)^{-\kappa}V_{\text{inf}}(\phi_\alpha)+e^K \tilde{V}(T,\phi_\alpha),
\ee
where
\begin{equation}
\begin{split}
V_{\text{mod}}&=e^{K_{\text{mod}}}\left[ K^{T\bar{T}}|D_T W_{\text{mod}}|^2 - 3 |W_{\text{mod}}|^2\right]\\[4pt]
V_{\text{inf}}&=e^{K_{\text{inf}}}\left[ K^{\alpha\bar{\alpha}}|D_\alpha W_{\text{inf}}|^2 - 3 |W_{\text{inf}}|^2\right].
\end{split}
\end{equation}
 In this case the potential is minimized at
$\tau_0+\delta T$ due to the effect of the inflationary large positive energy density where $\tau_0$ corresponding the SUSY Minkowski minimum if the scenario contain moduli only.
Since the modulus is very heavy it stabilizes quickly to its minimum and the inflationary potential get corrections after setting $T$ to its minimum as follows \cite{Buchmuller:2014vda}
\bea
V &=& (2\tau_0)^{-3}V_{\text{inf}}- \frac{\kappa}{2(2\tau_0)^{3\kappa/3}m_T}\Bigg\{ W_{\text{inf}}\Big[ V_{\text{inf}}
+ e^{K_{\text{inf}}}  K^{\alpha\bar{\alpha}}\partial_\alpha W_{\text{inf}}D_{\bar{\alpha}} \overline{W}_{\text{inf}} \Big] + h.c. \Bigg\}\n
\quad\quad  &-& \frac{\kappa e^{K_{\text{inf}}}}{(2\tau_0)^{2\kappa}m_T^2}\left| K^{\alpha\bar{\alpha}}D_\alpha W_{\text{inf}}\partial_{\bar{\alpha}} \overline{W}_{\text{inf}} \right|
\eea

Now consider a case of small field inflation such as the F-term Hybrid inflation \cite{Dvali:1994ms}, where the K\"ahler potential and superpotential are given by
\bea
K_{\text{inf}}&=&|\varphi|^2+ |\phi_1|^2+|\phi_2|^2, \\
W& =& \lambda\varphi\left(M^2 - \phi_1\phi_2\right),
\eea
where $\varphi$ is the inflaton and $\phi_1,\phi_2$ correspond to the waterfall fields. The hybrid inflation scenario supposes that for values of $\varphi > \varphi_c =M$, the potential is minimized at $\phi_1=\phi_2=0$ and therefor effectively the
$W_{\text{inf}}$ is given by
\be
W_{\text{inf}}=\lambda M^2 \varphi.
\ee
Accordingly, the corrected inflationary potential due to the moduli backreaction will contain a linear term in the inflaton \cite{Buchmuller:2014vda}
\be
V_{\text{inf}}(\varphi)= V_0\left[1-\frac{2\kappa \sqrt{V_0}}{m_T} {\text{Re}}(\varphi)\right]+\Delta V_1+V_{\text{sug}},
\ee
where $\Delta V_1$ is the one-loop correction to the potential and $V_{\text{sug}}$ are contributions due to supergravity which can be neglected. Here $V_0$
is the rescaled vacuum energy during inflation given by $V_0=\tilde{\lambda}^2 M^4$, with $\tilde{\lambda}^2=\frac{{\lambda}^2}{(2\tau_0)^{\kappa}}$. In this case the value of the spectral index can be improved $n_s=0.96$  compared to its value in the pure hybrid scenarios $n_s=0.98$ and so we don't need to go to non-minimal K\"ahler scenario \cite{Dvali:1994ms}.

Another example is the large field inflation  which  was successfully embedded in supergravity. In this case, a shift symmetry is used to avoid the $\eta$-problem as well as considering
stabilizer field $S$  \cite{Kawasaki:2000yn}. In this regard consider a form of the  K\"ahler potential and superpotential  given by
\bea
K_{\text{inf}} &=&-{1\over 2}|\varphi-\bar{\varphi}|^2+ |S|^2- \zeta|S|^4, \\
W_{\text{inf}} &=& Sf(\varphi),
\eea
where the quartic term in K\"ahler potential causes the stabilizer $S$ to get mass much larger than the Hubble parameter and so stabilizes quickly to the origin.
The K\"ahler potential is independent of the real part  ${\text{Re}}(\varphi)=\varphi_r$ due to the shift symmetry and therefor it will correspond to the inflaton.
The $S$ field and the imaginary part of $\varphi$ will stabilize at zero and hence the inflationary potential (which is given in pure inflation
scenario case by $V(\varphi_r)=|f(\varphi_r)|^2$) is corrected as follows \cite{Buchmuller:2014vda}
\be
V(\varphi_r)=\tilde{V}  \left[1-\kappa \frac{\tilde{V}}{m_T^2}- \kappa^2\frac{\tilde{V}^2}{m_T^2m_S^2} \right],
\ee
where $\tilde{V}(\varphi_r)= \frac{|f(\varphi_r)|^2}{(2 \tau_0)^\kappa}$. For $f(\varphi)=m \varphi$, we get the case chaotic inflation. It is worth noting that the
leading correction term here is proportional to ${1 \over m_T^2}$ whereas in the case of hybrid inflation it is ${1 \over m_T}$.

\section{Conclusions}
%
In this mini-review we have analyzed the problem of moduli stabilization in type IIB string theory with positive vacuum energy. We focused on KKLT and large volume scenarios, where geometrical fluxes and non-perturbative superpotentials  are required to stabilize complex structure moduli, dilation and K\"ahler moduli. We also discussed some possible mechanisms for uplifting the AdS minimum and making it a metastable de Sitter ground state. We have derived the soft SUSY breaking terms in these models. We showed that in KKLT these terms are not consistent with electroweak breaking conditions and hence they are not phenomenologically viable. While, in LVS we found that the scalar masses, gaugino masses and trilinear terms are universal and are given in terms of gravitino mass $m_{3/2}$.  We emphasized that a very heavy spectrum with $m_{3/2}\sim 1.5$ TeV is required to account for the lightest Higgs mass limit. However, in this case the relic abundance of the lightest  neutralino is not consistent with the measured limits. We also studied inflation scenarios associated with the moduli stabilization. We considered two examples of racetrack and K\"ahler inflation. Finally, we commented on the problem of moduli destabilization and moduli backreaction effects on inflation.

%
\section{Acknowledgments}
%
This work was partially supported by the STDF project 13858 and the ICTP grant AC-80.
A. M. would like to thank W. Abdallah for useful discussions.
%

\end{document}